# Remote magnon-magnon entanglement enhanced by squeezed-field interference

Yuan Gong,[1] Yan-Xue Cheng,[1] Wei Xiong,[1,2,*] and Jiaojiao Chen[1,†]
[1]*Department of Physics, Wenzhou University, Zhejiang 325035, China*
[2]*International Quantum Academy, Shenzhen 518048, China*

(Received 18 November 2025; accepted 13 May 2026; published 26 May 2026)

Cavity magnonics, owing to its strong magnon-photon coupling and excellent tunability, has attracted significant interest in quantum information science. However, achieving strong and robust macroscopic entanglement remains a long-standing challenge due to the inherently linear nature of the beam-splitter interaction. Here, we propose an experimentally feasible scheme to generate and enhance macroscopic entanglement between two remote magnon modes by injecting squeezed vacuum fields (SVFs) into coupled microwave cavities. We demonstrate that even a single SVF applied to one cavity can induce steady magnon-magnon entanglement, while applying two SVFs (the double-squeezed configuration) enables selective activation of two independent entanglement channels associated with the cavity supermodes. Remarkably, quantum interference between the two SVFs allows for phase-controlled enhancement of entanglement, resulting in significantly improved robustness against cavity dissipation and thermal noise. Under realistic parameters, the survival temperature of quantum entanglement increases from approximately 260 to 450 mK. Our results establish a versatile and controllable approach to generating and enhancing quantum entanglement through quantum interference between squeezed vacuum fields.

## I. INTRODUCTION

Magnons, the quanta of spin waves, are collective excitations of spin ensembles in magnetically ordered materials [1,2]. Thanks to their high spin density and low damping [3], magnons in yttrium-iron-garnet (YIG) spheres have attracted considerable interest, both theoretically and experimentally, in quantum information science and spintronics [4,5]. Benefiting from these unique properties, magnons can be coupled to diverse quantum systems, such as microwave photons [6], superconducting qubits [7], and phonons [8]. As a result, versatile hybrid quantum systems are constructed. As a typical magnon-based hybrid system, cavity magnonics [4] is established by magnons strongly coupled to microwave photons in a microwave cavity and has witnessed remarkable progress over the past decade, including dark modes [9], non-Hermitian dynamics [10–15], spin current [16], level attraction [17,18], near-perfect absorption [19], and quantum magnon [20,21]. It is evident that these phenomena are beneficial to the linear beam-splitter magnon-photon interaction in cavity magnonics. However, relying solely on such a coupling mechanism is extremely unfavorable for realizing macroscopic quantum entanglement [22], which is essential not only for quantum information processing [23] but also for probing the boundary between classical and quantum physics [24]. Therefore, generating macroscopic quantum entanglement in cavity magnonics remains a long-standing challenge.

*Contact author: xiongweiphys@wzu.edu.cn
†Contact author: jjchenphys@wzu.edu.cn

To address this challenge, one approach is to couple magnons to phonons via the nonlinear magnetostrictive interaction [25–28], analogous to the optomechanical coupling in cavity optomechanics [29]. In the linearized regime, the two-mode squeezing terms generate magnon-phonon entanglement, while the beam-splitter terms simultaneously cool the phonons to facilitate the observation of entanglement. Assisted by the beam-splitter magnon-photon interaction, the magnon-phonon entanglement can be transferred among photons, phonons, and magnons, enabling the preparation of arbitrary bipartite or tripartite entanglement under appropriate conditions [25]. An alternative strategy is to directly couple magnons to microwave photons in electromechanical systems, effectively implementing cavity-magnon optomechanics in the microwave regime [30,31]. In such systems, macroscopic entanglement arises from radiation-pressure-like interactions between photons and phonons, a mechanism closely analogous to that in cavity magnomechanics. Another promising route is to exploit the intrinsic Kerr nonlinearity of magnons in YIG spheres [32], which originates from magnetocrystalline anisotropy [33] and has been demonstrated experimentally [34,35]. Beyond quantum entanglement, the magnon Kerr nonlinearity can also be used to investigate other rich physics, including bistability [36] and multistability [37], magnon-mediated remote spin-spin strong coupling [38], spin-polariton [39] and phonon-polariton strong coupling [40], exponentially enhanced tripartite interactions [41], quantum phase transitions [42–44], nonreciprocal entanglement [30,45,46], nonreciprocal photon blockade [47], ground-state cooling [48], and ultrasensitive detection [49]. A further approach is to inject squeezed vacuum fields (SVFs) into cavities to squeeze photons [50–55]. The quantum correlations of the squeezed fields can then be transferred to magnons via the beam-splitter magnon-photon interaction. As a result, two nonlocal magnon modes in a cavity [53,54], magnons in uncoupled double cavities [52], and arbitrary pairs of remote magnon modes in a chain of cavity magnonics [55] can be entangled.

Building upon the SVF approach, we propose an experimentally feasible scheme to not only generate but also enhance entanglement between two remote magnon modes. The system comprises two remotely coupled cavity modes connected by a coaxial cable [56], with each cavity mode coupled to a local magnon mode in a YIG sphere. We first show that, in the single-squeezed configuration (i.e., when only one SVF is applied), magnon-magnon entanglement can be generated exclusively through a specific supermode that is resonant with the SVF, implying that only a single entanglement channel is effectively activated. The two resulting supermodes arise from the hybridization of two strongly coupled cavities. In the supermode picture, the magnon-cavity interactions are effectively mapped onto magnon-supermode interactions, and an SVF applied to a cavity is equivalent to applying SVFs to the supermodes. Under the resonant condition, the quantum correlations carried by the SVF are efficiently transferred to the magnon modes through the effective magnon-supermode coupling, thereby generating entanglement between the two remote magnon modes. We then demonstrate that, in the double-squeezed configuration (i.e., when two SVFs are, respectively, applied to the two cavity modes), magnon-magnon entanglement can be equivalently realized via either of the two supermodes, indicating that two entanglement channels can be selectively activated. Moreover, the magnon-magnon entanglement in the double-squeezed configuration is significantly enhanced compared with the single-squeezed case, owing to the occurrence of quantum interference between the two SVFs at the supermode interface. This interference provides an additional route for coherent control of magnon-magnon entanglement through the squeezing phases. Furthermore, the magnon-magnon entanglement exhibits enhanced robustness against cavity dissipation and elevated bath temperatures in the double-squeezed configuration. Under realistic parameters, the survival temperature for quantum entanglement increases from approximately 260 mK in the single-squeezed configuration to about 450 mK in the double-squeezed case.

Note that our work differs substantially from previous studies [52–55]. In Ref. [52], a two-mode SVF is employed to entangle two remote magnon modes located in two uncoupled cavities. In Refs. [53,54], both magnon modes are hosted within a single cavity, where magnon-magnon entanglement is generated by applying a single SVF; consequently, neither cavity supermodes nor quantum interference effects are involved. In Ref. [55], an array of coupled cavities is considered, each containing a magnon mode. By applying a single SVF to the central magnon mode, bipartite and tripartite entanglement among magnons is achieved. However, generating an SVF directly for magnon modes remains experimentally challenging with current technology. Moreover, although coupled cavities are included in that work, they are adiabatically eliminated to induce effective magnon-magnon couplings. In contrast, our work focuses on how quantum interference between two SVFs influences magnon-magnon entanglement. Specifically, (1) the magnon-magnon entanglement is phase controlled. In particular, when the two SVFs are phase matched, the magnon-magnon entanglement is significantly enhanced; (2) the resulting entanglement exhibits increased robustness against cavity decay and elevated bath temperatures. In addition, (3) we explicitly reveal the distinct roles played by the two cavity supermodes in generating magnon-magnon entanglement under both the single- and double-squeezed configurations.

The rest of this paper is organized as follows. In Sec. II, we give a brief introduction of the proposed system and the corresponding Hamiltonian and its dynamics are given. Then, we show how to generate the magnon-magnon entanglement with a single SVF and enhance the entanglement with double SVFs in Sec. III. Finally, a summary is concluded in Sec. IV.

## II. MODEL AND QUANTUM LANGEVIN EQUATION

### A. Model

We consider a coupled cavity-magnon system, as illustrated in Fig. 1, composed of two YIG spheres and two microwave cavities connected by a coaxial cable. In each cavity, a YIG sphere is biased by a uniform static magnetic field to excite a magnon mode and is positioned near the region of maximum magnetic field to achieve strong coupling between the magnon mode and the cavity mode. Under the rotating-wave approximation, the total Hamiltonian can be written as (with $\hbar = 1$)

$$H = H_0 + \sum_{j=1}^{2} g_j(a_j m_j^\dagger + a_j^\dagger m_j) + J(a_1 a_2^\dagger + a_1^\dagger a_2), \tag{1}$$

where $H_0 = \sum_{j=1,2}(\omega_{a_j} a_j^\dagger a_j + \omega_{m_j} m_j^\dagger m_j)$ is the free Hamiltonian, with $\omega_{a_j}$ and $\omega_{m_j}$ representing the resonance frequencies of the $j$th cavity and magnon modes, respectively. Here, $a_j$ ($a_j^\dagger$) and $m_j$ ($m_j^\dagger$) are the annihilation (creation) operators of the $j$th cavity and magnon modes. The parameter $g_j$ denotes the coupling strength between the $j$th cavity mode and its embedded magnon mode, and the intercavity coupling strength is characterized by $J$. In our proposal, we assume that the two cavity modes are strongly coupled, which results in their hybridization into two supermodes defined as $a_\pm = (a_1 \pm a_2)/\sqrt{2}$. In the supermode basis, the subsystem composed of the two coupled cavities can be diagonalized as

$$\sum_{j=1,2} \omega_{a_j} a_j^\dagger a_j + J(a_1^\dagger a_2 + a_1 a_2^\dagger) \rightarrow \omega_+ a_+^\dagger a_+ + \omega_- a_-^\dagger a_-,$$

where $\omega_\pm = \frac{1}{2}(\omega_{a_1} + \omega_{a_2} \pm \sqrt{(\omega_{a_1} - \omega_{a_2})^2 + 4J^2})$. For two resonant cavities ($\omega_{a_1} = \omega_{a_2} = \omega_a$), the supermode frequencies reduce to $\omega_\pm = \omega_a \pm J$, yielding two resonant peaks separated by $\Delta\omega = 2J$, as illustrated in Fig. 1(b). In terms of the supermodes $a_\pm$, the annihilation operators of the original cavity modes can be written as $a_1 = (a_+ + a_-)/\sqrt{2}$ and $a_2 = (a_+ - a_-)/\sqrt{2}$. Consequently, the Hamiltonian in Eq. (1) can be rewritten as

$$H' = H_0' + (M_+ a_+^\dagger + M_- a_-^\dagger + \text{H.c.}),$$

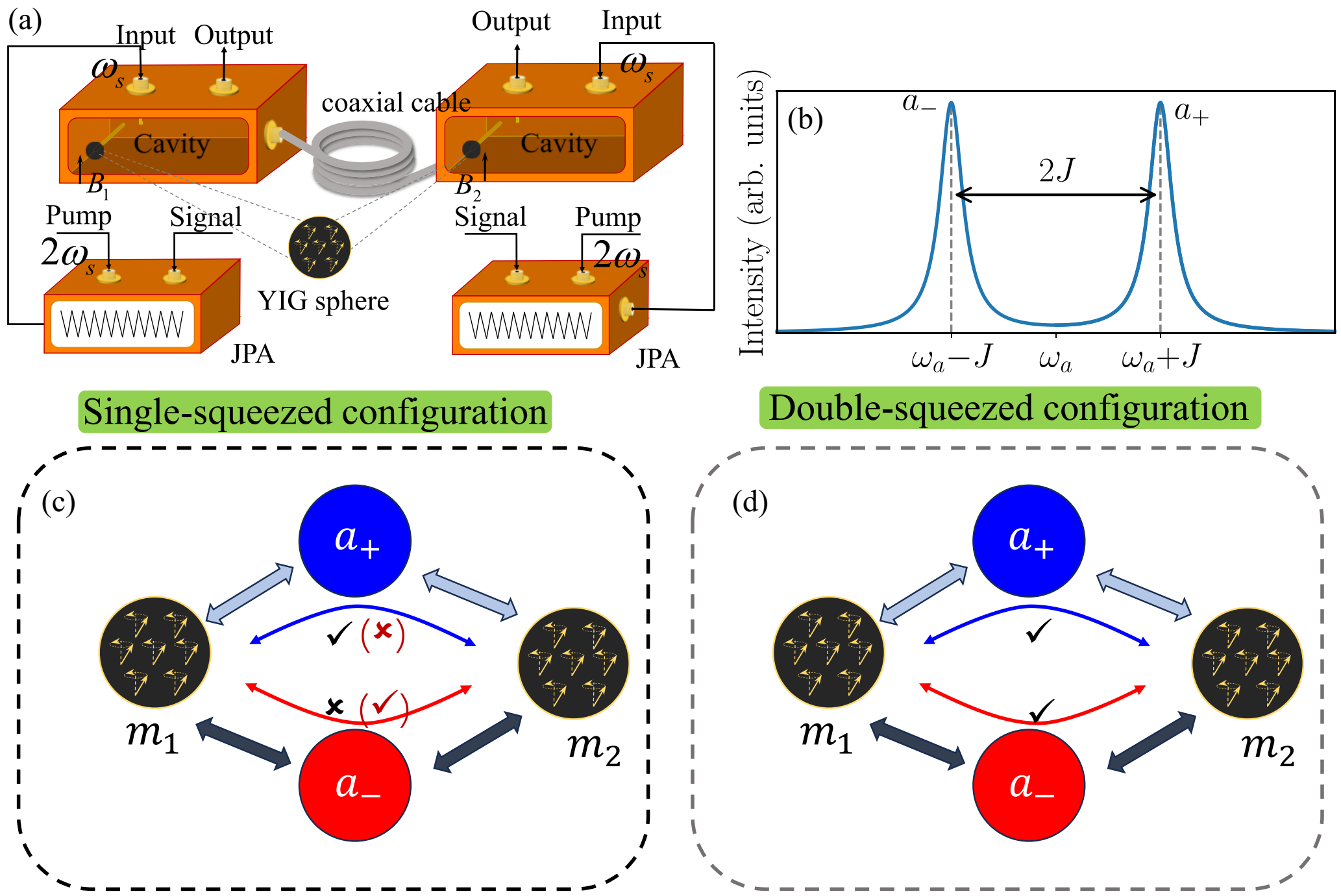

FIG. 1. (a) Schematic diagram of the coupled cavity-magnon system. Two microwave cavities, connected by a coaxial cable, are, respectively, driven by two single-mode squeezed vacuum fields with frequency $\omega_s$, generated by two Josephson parametric amplifiers (JPAs) pumped at frequency $2\omega_s$. (b) Two supermodes with frequencies $\omega_a \pm J$, formed by two resonant cavities, are depicted. $2J$ is the separation between two supermodes. (c) The mechanism of the generation of the magnon-magnon entanglement in the single-squeezed configuration, where only one *determined* supermode is resonant with the injected squeezed field. (d) The mechanism of the generation of the magnon-magnon entanglement in the double-squeezed configuration, where two supermode can be *selectively* resonant with the injected squeezed fields, via tuning the detunings of the cavity.

where $H_0' = \omega_+ a_+^\dagger a_+ + \omega_- a_-^\dagger a_- + \omega_{m_1} m_1^\dagger m_1 + \omega_{m_2} m_2^\dagger m_2$ and $M_\pm = (g_1 m_1 \pm g_2 m_2)/\sqrt{2}$. This form of the Hamiltonian shows that the two magnon modes are effectively mediated by the two cavity supermodes, which plays a crucial role in generating magnon-magnon entanglement. The underlying mechanism is illustrated in Figs. 1(c) and 1(d).

### B. Quantum Langevin equation

Taking into account the dissipation of both the cavity and magnon modes, the system dynamics governed by the Hamiltonian in Eq. (1) can be described by the following quantum Langevin equations (QLEs) [57]:

$$\begin{aligned}
\dot{a}_1 &= -(\kappa_{a_1} + i\omega_{a_1})a_1 - ig_1 m_1 - iJa_2 + \sqrt{2\kappa_{a_1}}a_1^{\mathrm{in}},\\
\dot{a}_2 &= -(\kappa_{a_2} + i\omega_{a_2})a_2 - ig_2 m_2 - iJa_1 + \sqrt{2\kappa_{a_2}}a_2^{\mathrm{in}},\\
\dot{m}_j &= -(\kappa_{m_j} + i\omega_{m_j})m_j - ig_j a_j + \sqrt{2\kappa_{m_j}}m_j^{\mathrm{in}},
\end{aligned} \tag{2}$$

where $\kappa_{a_j}$ and $\kappa_{m_j}$ denote the decay rates of the $j$th cavity and magnon modes, respectively. The operators $a_j^{\mathrm{in}}$ and $m_j^{\mathrm{in}}$ represent the input noise operators for the $j$th cavity and magnon modes.

These input noise operators have zero mean values and satisfy the correlation relations [57]:

$$\begin{aligned}\langle O_j^{\text{in},\dagger}(t)O_j^{\text{in}}(t')\rangle &= N_{O_j}\delta(t-t'), \quad O=a,m,\\ \langle O_j^{\text{in}}(t)O_j^{\text{in},\dagger}(t')\rangle &= (N_{O_j}+1)\delta(t-t'),\end{aligned} \tag{3}$$

where $N_{O_j}=[\exp(\hbar\omega_{O_j}/k_BT)-1]^{-1}$ is the mean thermal occupation number, with $T$ being the bath temperature and $k_B$ the Boltzmann constant.

It is well known that the system with only beam-splitter interactions and input noise described by Eq. (3) cannot generate macroscopic entanglement [22]. To overcome this limitation, each cavity mode is driven by a single-mode SVF instead of the original thermal noise (see Fig. 1), which can be readily generated via degenerate parametric down-conversion in flux-driven Josephson parametric amplifiers [58–67]. Experimentally, it has been demonstrated that JPAs can achieve squeezing levels up to 10 dB [62]. When a JPA is pumped at frequency $2\omega_s$ while its signal input port is kept in vacuum, it generates a squeezed microwave vacuum field centered at $\omega_s$ (see Fig. 1) [63]. Alternatively, a squeezed vacuum field can also be realized using a degenerate parametric amplifier driven by a microwave pump [68]. In this double-squeezed configuration, the QLEs in Eq. (2) are modified as

$$\begin{aligned}\dot a_1 &= -(\kappa_{a_1}+i\Delta_{a_1})a_1 - ig_1m_1 - iJa_2 + \sqrt{2\kappa_{a_1}}s_1^{\text{in}},\\ \dot a_2 &= -(\kappa_{a_2}+i\Delta_{a_2})a_2 - ig_2m_2 - iJa_1 + \sqrt{2\kappa_{a_2}}s_2^{\text{in}},\\ \dot m_j &= -(\kappa_{m_j}+i\Delta_{m_j})m_j - ig_ja_j + \sqrt{2\kappa_{m_j}}m_j^{\text{in}},\end{aligned} \tag{4}$$

where $\Delta_{a_j}=\omega_{a_j}-\omega_s$ and $\Delta_{m_j}=\omega_{m_j}-\omega_s$ are the detunings of the cavity and magnon modes from the SVF frequency $\omega_s$. The operators $s_j^{\text{in}}$ denote the single-mode SVF acting on the $j$th cavity mode, satisfying [69]

$$\begin{aligned}\langle s_j^{\text{in},\dagger}(t)s_j^{\text{in}}(t')\rangle &= N_{s_j}\delta(t-t'),\\ \langle s_j^{\text{in}}(t)s_j^{\text{in},\dagger}(t')\rangle &= (N_{s_j}+1)\delta(t-t'),\\ \langle s_j^{\text{in}}(t)s_j^{\text{in}}(t')\rangle &= M_{s_j}\delta(t-t'),\\ \langle s_j^{\text{in},\dagger}(t)s_j^{\text{in},\dagger}(t')\rangle &= M_{s_j}^*\delta(t-t'),\end{aligned} \tag{5}$$

where $N_{s_j}=\sinh^2 r_j$ and $M_{s_j}=e^{i\theta_j}\sinh r_j\cosh r_j$, with $r_j$ and $\theta_j$ denoting the squeezing parameter and phase of the SVF acting on the $j$th cavity mode.

### C. Covariance matrix and logarithmic negativity

By introducing the quadrature operators for the cavity and magnon modes, $x_{O_j}=(O_j^\dagger+O_j)/\sqrt{2}$, $y_{O_j}=i(O_j^\dagger-O_j)/\sqrt{2}$, Eq. (4) can be rewritten in the matrix form as

$$\dot u(t) = Au(t) + n(t), \tag{6}$$

where $u(t)=[x_{a_1},y_{a_1},x_{a_2},y_{a_2},x_{m_1},y_{m_1},x_{m_2},y_{m_2}]^T$, $A$ is the drift matrix,

$$A=\begin{pmatrix} -\kappa_{a_1} & \Delta_{a_1} & 0 & J & 0 & g_1 & 0 & 0\\ -\Delta_{a_1} & -\kappa_{a_1} & -J & 0 & -g_1 & 0 & 0 & 0\\ 0 & J & -\kappa_{a_2} & \Delta_{a_2} & 0 & 0 & 0 & g_2\\ -J & 0 & -\Delta_{a_2} & -\kappa_{a_2} & 0 & 0 & -g_2 & 0\\ 0 & g_1 & 0 & 0 & -\kappa_{m_1} & \Delta_{m_1} & 0 & 0\\ -g_1 & 0 & 0 & 0 & -\Delta_{m_1} & -\kappa_{m_1} & 0 & 0\\ 0 & 0 & 0 & g_2 & 0 & 0 & -\kappa_{m_2} & \Delta_{m_2}\\ 0 & 0 & -g_2 & 0 & 0 & 0 & -\Delta_{m_2} & -\kappa_{m_2}\end{pmatrix}, \tag{7}$$

and $n(t) = [\sqrt{2\kappa_{a_1}}x_{s_1}^{\rm in}(t), \sqrt{2\kappa_{a_1}}y_{s_1}^{\rm in}(t), \sqrt{2\kappa_{a_2}}x_{s_2}^{\rm in}(t), \sqrt{2\kappa_{a_2}}y_{s_2}^{\rm in}(t), \sqrt{2\kappa_{m_1}}x_{m_1}^{\rm in}(t), \sqrt{2\kappa_{m_1}}y_{m_1}^{\rm in}(t), \sqrt{2\kappa_{m_2}}x_{m_2}^{\rm in}(t), \sqrt{2\kappa_{m_2}}y_{m_2}^{\rm in}(t)]^T$ is the vector of input noises. To ensure system stability, the Routh-Hurwitz criterion must be satisfied; namely, all eigenvalues of the drift matrix $A$ must have negative real parts. This condition is numerically verified throughout our subsequent analysis. Since the system dynamics are linear and the input noises are Gaussian, any Gaussian input state remains Gaussian, and the steady state of the system is therefore a continuous-variable four-mode Gaussian state, fully characterized by the $8 \times 8$ covariance matrix $V$, defined as $V_{pq}(t) = \frac{1}{2}\langle u_p(t)u_q(t') + u_q(t')u_p(t)\rangle$ $(p, q = 1, 2, \ldots, 8)$.

In the long-time limit ($t \to \infty$), the system reaches a stable steady state, and the covariance matrix $V$ can be obtained by solving the Lyapunov equation [29,70]

$$AV + VA^T = -D, \tag{8}$$

where $D$ is the diffusion matrix, defined via

$$\tfrac{1}{2}\langle n_p(t)n_q(t') + n_q(t')n_p(t)\rangle = D_{pq}\delta(t - t'). \tag{9}$$

Specifically, it can be written as a direct sum

$$D = D_{s_1} \oplus D_{s_2} \oplus D_{m_1} \oplus D_{m_2}, \tag{10}$$

with

$$D_{s_j} = \kappa_{a_j}\begin{pmatrix} 2N_{s_j} + 1 + 2{\rm Re}[M_{s_j}] & 2{\rm Im}[M_{s_j}] \\ 2{\rm Im}[M_{s_j}] & 2N_{s_j} + 1 - 2{\rm Re}[M_{s_j}] \end{pmatrix} \tag{11}$$

and

$$D_{m_j} = \kappa_{m_j}\,{\rm Diag}[2N_{m_j} + 1, 2N_{m_j} + 1]. \tag{12}$$

Once the covariance matrix $V$ is obtained, the bipartite entanglement between the two magnon modes can be quantified using the *logarithmic negativity* [71,72]

$$E_N \equiv \max[0, -\ln 2\eta^-], \tag{13}$$

with

$$\eta^- = 2^{-1/2}\left[\Sigma - (\Sigma^2 - 4\det V_{mm})^{1/2}\right]^{1/2}, \tag{14}$$

where $\Sigma = \det A + \det B - 2\det C$ and

$$V_{mm} = \begin{pmatrix} A & C \\ C^T & B \end{pmatrix} \tag{15}$$

is the $4 \times 4$ covariance submatrix associated with the two magnon modes [73]. Here, $A$, $B$, and $C$ are the $2 \times 2$ blocks of $V_{mm}$. A positive logarithmic negativity, i.e., $E_N > 0$, indicates the presence of bipartite entanglement between the two magnon modes.

## III. QUANTUM-INTERFERENCE-ENHANCED MAGNON-MAGNON ENTANGLEMENT

### A. Single-squeezed configuration

We first demonstrate that magnon-magnon entanglement can be generated by driving only a single cavity mode with a single-mode SVF. Without loss of generality, we assume that the SVF is injected into cavity mode $a_1$. In this single-squeezed configuration, the input noise term $s_2^{\rm in}$ in Eq. (4) reduces to the vacuum noise operator $a_2^{\rm in}$ from Eq. (2), and its correlation function reverts to the form given in Eq. (3). Consequently, the diffusion matrix $D$ in Eq. (10) becomes

$$D = D_{s_1} \oplus D_{a_2} \oplus D_{m_1} \oplus D_{m_2}, \tag{16}$$

where $D_{a_2} = \kappa_{a_2} \times {\rm Diag}[2N_{a_2} + 1, 2N_{a_2} + 1]$.

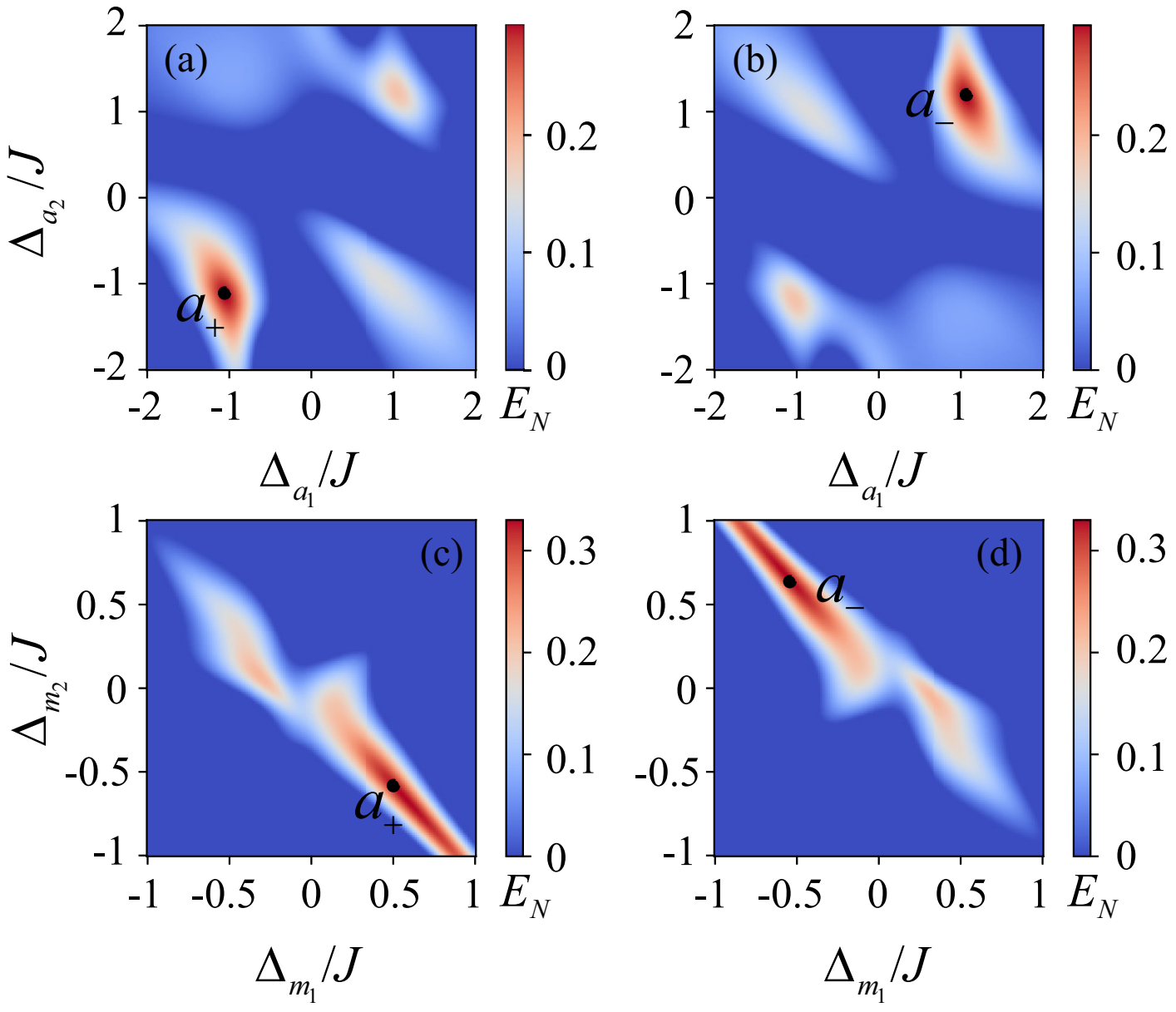

FIG. 2. Logarithmic negativity $E_N$ of magnon-magnon entanglement vs (a), (b) normalized cavity detunings ($\Delta_{a_1}$, $\Delta_{a_2}$) and (c), (d) normalized magnon detunings ($\Delta_{m_1}$, $\Delta_{m_2}$) in the single-squeezed configuration. In panels (a) and (b), we, respectively, take ($\Delta_{m_1}$, $\Delta_{m_2}$) = $(0.5J, -0.5J)$ and $(-0.5J, 0.5J)$. In panels (c) and (d), $\Delta_{a_1} = \Delta_{a_2} = -J$ and $+J$ are fixed. The numerical results reveal that a specific supermode ($a_+$ or $a_-$) is resonant with the squeezed field to achieve optimal magnon-magnon entanglement.

By numerically solving Eq. (8) in the single-squeezed configuration, we plot magnon-magnon entanglement versus the normalized cavity and magnon detunings in Fig. 2. The results show that the squeezed field can only be resonant with one specific supermode ($a+$ or $a_-$) at fixed cavity or magnon detunings to achieve optimal entanglement. This indicates that only one entanglement channel is activated to generate magnon-magnon entanglement, as illustrated in Fig. 1(c). Specifically, when $\Delta_{m_1} = 0.5J$ and $\Delta_{m_2} = -0.5J$, the optimal entanglement is achieved where the squeezed field is resonant with the upper supermode $a_+$ with frequency $\omega_+ = \omega_a + J$ [see Fig. 2(a)]. But when $\Delta_{m_1} = -0.5J$ and $\Delta_{m_2} = 0.5J$, the optimal entanglement is achieved where the squeezed field is resonant with the lower supermode $a_-$ with frequency $\omega_- = \omega_a - J$ [see Fig. 2(b)]. The dependence of entanglement on magnon detunings is further illustrated in Figs. 2(c) and 2(d), which correspond to mediation through the upper and lower supermodes, respectively. Clearly, the entanglement degrades as the magnon detunings deviate from their optimal values. In Fig. 2, the following experimental parameters are taken [74]: $\kappa_{a_1}/2\pi = \kappa_{a_2}/2\pi = \kappa_a/2\pi = 5$ MHz, $\kappa_{m_1} = \kappa_{m_2} = \kappa_a/5$, $J = 4\kappa_a$, $g_1 = g_2 = 2\kappa_a$, $r_1 = r_2 = 0.9$, and $\theta_1 = \theta_2 = 0$.

### B. Double-squeezed configuration

The magnon-magnon entanglement can be significantly enhanced by driving both cavity modes with SVFs, referred to as the double-squeezed configuration. Similar to the single-squeezed case, optimal entanglement is achieved when the SVFs are resonant with one of the supermodes, as shown in Figs. 3(a) and 3(b). Unlike the single-squeezed configuration, however, the two supermodes contribute equally to mediating entanglement. By appropriately tuning the cavity detunings, either supermode can be selectively resonant with the squeezed field to transfer quantum correlations from the SVFs to the magnon modes, as explicitly demonstrated in Figs. 3(c) and 3(d). This indicates that double entanglement channels can be selectively activated [see Fig. 1(d)]. The enhancement of the magnon-magnon entanglement arises because, although each SVF directly drives its corresponding cavity mode, the strong intercavity coupling allows the hybridized supermodes to be indirectly excited by both SVFs, giving rise to quantum interference at the interface of the resonant supermode.

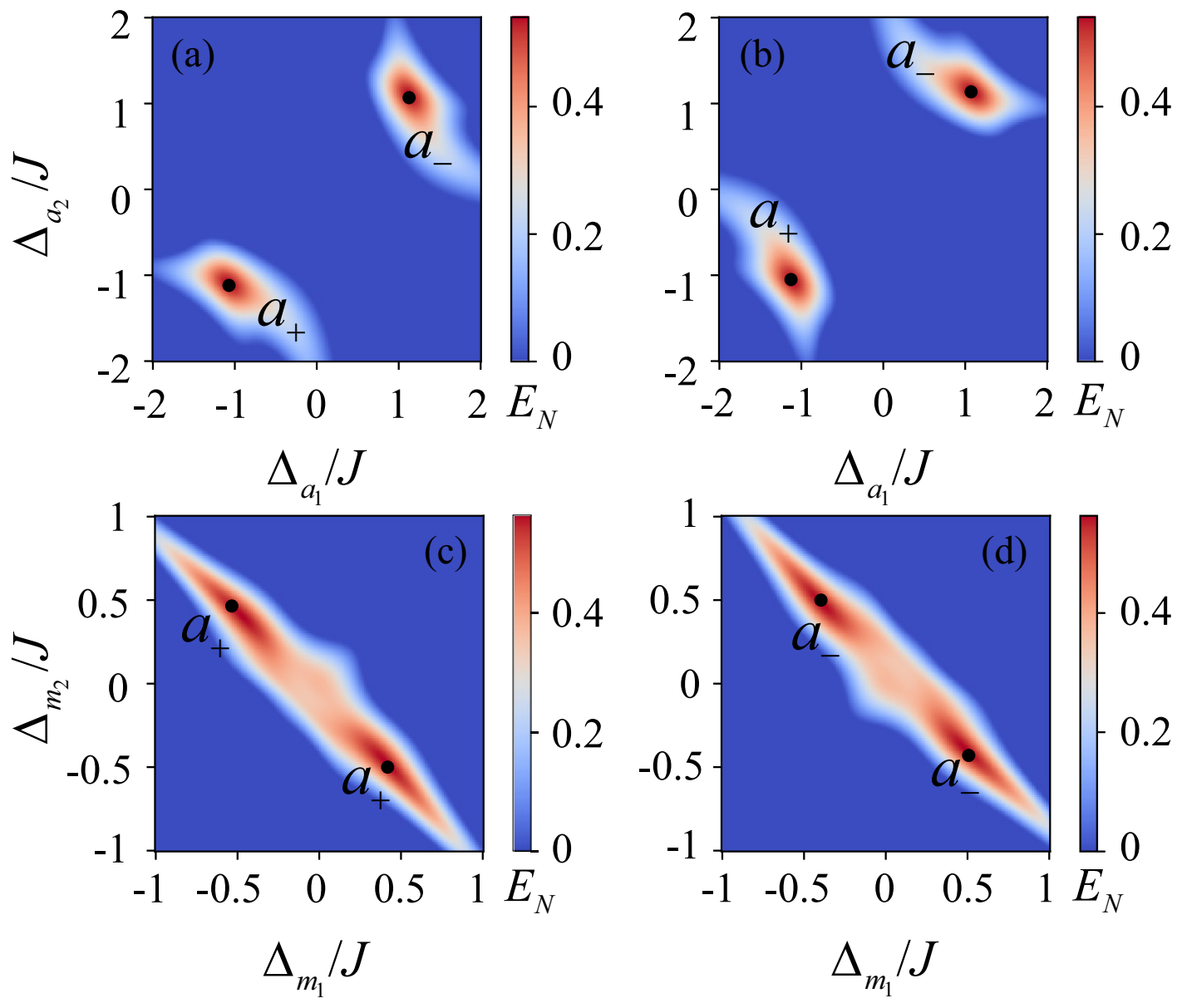

FIG. 3. Logarithmic negativity $E_N$ of magnon-magnon entanglement vs (a), (b) normalized cavity detunings ($\Delta_{a_1}$, $\Delta_{a_2}$) and (c), (d) normalized magnon detunings ($\Delta_{m_1}$, $\Delta_{m_2}$) in the double-squeezed configuration. The magnon detunings in panel (a) [panel (b)] are the same as that in Fig. 2(a) [Fig. 2(b)], and the cavity detunings in panel (c) [panel (d)] are identical with that in Fig. 2(c) [Fig. 2(d)]. The numerical results show that two supermodes ($a_+$ and $a_-$) can be selectively resonant with the squeezed fields to achieve optimal magnon-magnon entanglement.

The interference effect is clearly seen in Fig. 4(a), where the upper supermode is resonant with the squeezed field as an example. The magnon-magnon entanglement can be tuned by the relative phases of the SVFs in the double-squeezed configuration: It is constructively enhanced when the two SVFs are in phase and completely suppressed when they are out of phase. This behavior contrasts with the single-squeezed configuration, where the entanglement cannot be controlled by the local phase of a single SVF [Fig. 4(b)]. In both configurations, the entanglement can also be adjusted by the squeezing strength [Figs. 4(c) and 4(d)], indicating the existence of an optimal squeezing parameter for maximizing entanglement.

### C. Robustness and imperfections

Furthermore, the double-squeezed configuration exhibits superior robustness against cavity decay. Strong magnon-magnon entanglement can be preserved even for relatively large decay rates of cavity mode $a_2$ [Fig. 5(a)], whereas in the single-squeezed configuration significant entanglement requires extremely small decay rates [Fig. 5(b)], which poses substantial experimental challenges. As the decay rate increases, the entanglement decreases nonmonotonically in the double-squeezed configuration [Fig. 5(c)], but monotonically in the single-squeezed case [Fig. 5(d)], further confirming the enhanced robustness of the double-squeezed setup. The influence of bath temperature on the entanglement is further investigated in Fig. 6(a). The double-squeezed configuration (blue solid curve) exhibits significantly greater tolerance to thermal noise than the single-squeezed configuration. Under feasible parameters, the entanglement survives up to approximately 450 mK, whereas it vanishes around 260 mK in the single-squeezed case. Moreover, across the entire temperature range, the entanglement in the double-squeezed configuration remains consistently higher, confirming its superior robustness against thermal decoherence.

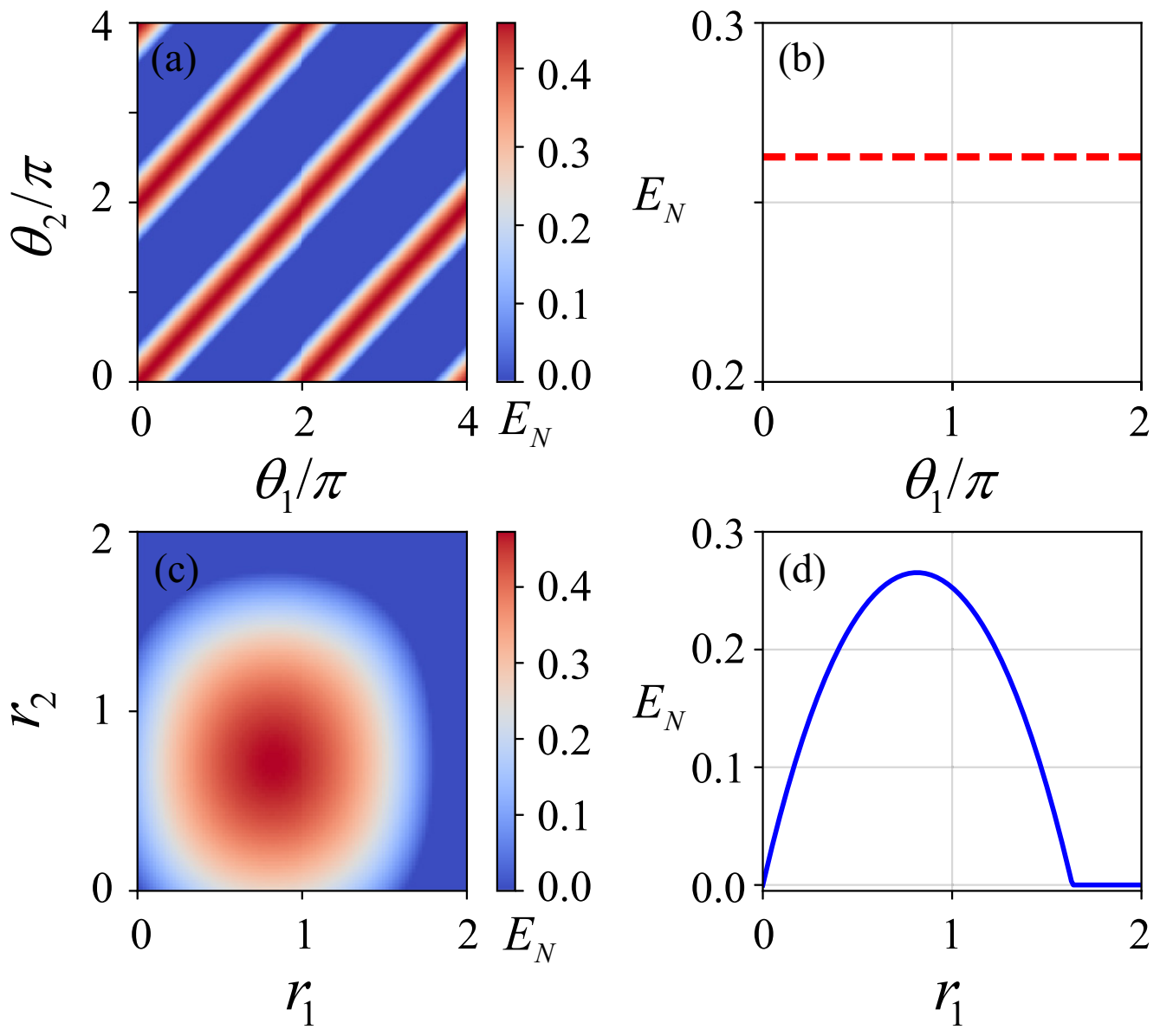

FIG. 4. Dependence of magnon-magnon entanglement in the double-squeezed configuration on (a) squeezing phases $(\theta_1, \theta_2)$ and (c) squeezing parameters $(r_1, r_2)$. The parameters $r_1 = r_2 = 0.9$ are used in panel (a), while $\theta_1 = \theta_2 = 0$ are fixed in panel (c). For comparison, the corresponding magnon-magnon entanglement in the single-squeezed configuration is shown in panels (b) and (d). The cavity and magnon detunings are set to $\Delta_{a_1} = \Delta_{a_2} = -J$ and $\Delta_{m_1} = -\Delta_{m_2} = 0.5J$, respectively. The results demonstrate that magnon-magnon entanglement can be tuned by the relative phase between the two squeezed fields and can be significantly enhanced via quantum interference between them.

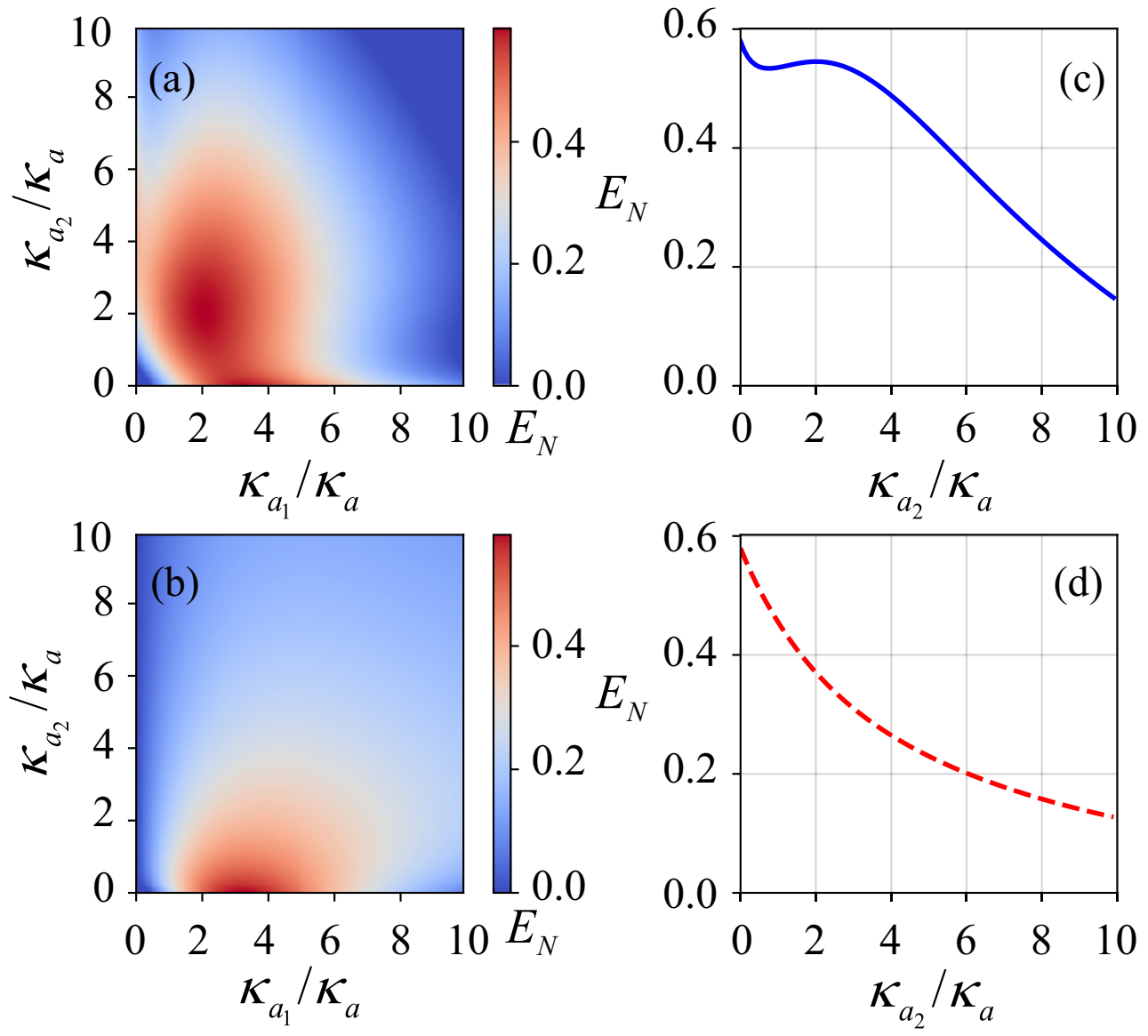

FIG. 5. Robustness of magnon-magnon entanglement against normalized cavity decays $(\kappa_1, \kappa_2)$ in the (a) double- and (b) single-squeezed configurations. The corresponding cross sections at $\kappa_{a_1} = 2.5\kappa_a$ are shown in panels (c) and (d). The cavity and magnon detunings are the same as in Fig. 4.

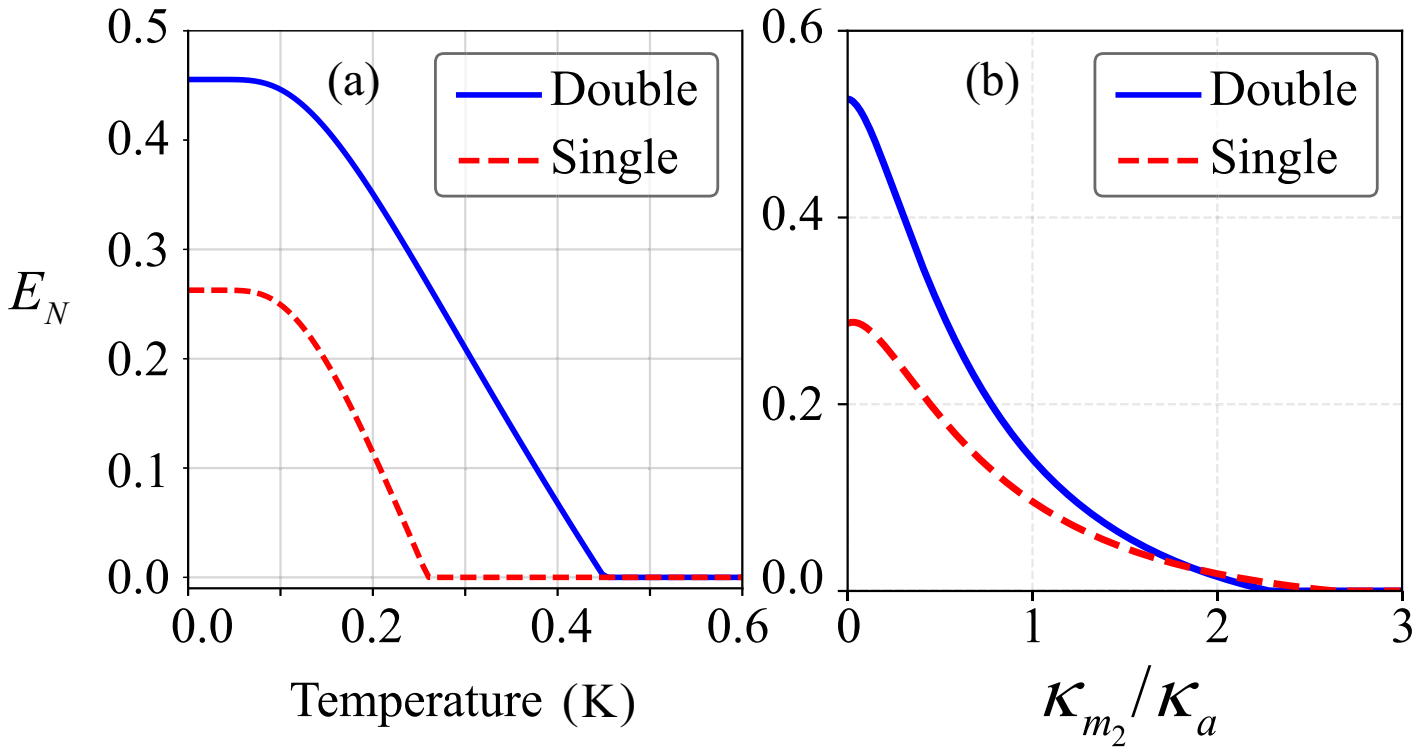

FIG. 6. (a) Temperature dependence of magnon-magnon entanglement for double-squeezed and single-squeezed configurations. (b) Effect of the magnon decay rate on entanglement in double-squeezed and single-squeezed configurations, with $\kappa_{m_1} = \kappa_a/5$ fixed. In both panels, the blue solid and red dashed curves represent the double-squeezed and single-squeezed configurations, respectively. The cavity and magnon detunings are identical to those in Fig. 4.

The enhanced robustness and thermal tolerance of the magnon-magnon entanglement in the double-squeezed configuration originate from a phase-sensitive quantum interference effect between the two squeezed fields. In the single-squeezed configuration, only one squeezed vacuum field is applied, so each supermode formed by the two strongly coupled cavity modes effectively experiences a single-squeezed field, and no quantum interference occurs. In contrast, in the double-squeezed configuration, each cavity mode is driven by its own squeezed vacuum field, such that each cavity supermode is effectively driven by two squeezed fields. Phase-sensitive quantum interference then occurs at the cavity supermode. Since squeezed fields can suppress the cavity linewidth and improve the effective cooling temperature [75], which is essential for the survival of quantum entanglement, the interference between the two squeezed fields further reduces the cavity linewidth and enhances the effective cooling, leading to the observed improvement in the robustness and thermal tolerance of macroscopic entanglement (Figs. 5 and 6).

Finally, several realistic imperfections should be taken into account. In our proposal, cable loss is neglected for simplicity; however, in realistic experiments such loss is unavoidable. Because the cable is directly connected to the cavity, cable loss effectively increases the decay rate of the cavity mode. As shown in Fig. 5, a larger cavity decay rate leads to reduced magnon-magnon entanglement, implying that cable loss degrades the generated entanglement. In addition, ideal magnon modes of the YIG sphere are assumed; i.e., magnon inhomogeneous broadening is not included. When this imperfection is taken into account, the increased magnon linewidth leads to reduced entanglement, as demonstrated in Fig. 6(b). Moreover, for simplicity we assume identical magnon-cavity coupling strengths in our proposal, which may be difficult to achieve in practical experiments. To investigate the influence of asymmetric couplings on the magnon-magnon entanglement, Fig. 7 is presented. When the two coupling strengths are nonidentical but comparable, the entanglement is only slightly reduced in both the double- and single-squeezed configurations [see panels (a) and (b), respectively]. However, when the couplings become strongly asymmetric, the magnon-magnon entanglement is severely degraded.

Besides, the enhancement of magnon-magnon entanglement in our double-squeezed scheme critically relies on quantum interference between the two squeezed vacuum fields, which requires maintaining phase stability between them. From an experimental perspective, stabilizing the rel-ative phase between multiple JPAs is an important requirement for implementing phase-sensitive interference of squeezed microwave fields. In circuit-QED experiments, this challenge is typically addressed by driving multiple JPAs with a common microwave reference, enabling phase-locked operation and long-term stabilization of the relative squeezing phase. Phase-sensitive interference and entanglement generation using squeezed microwave fields have already been experimentally demonstrated [64,76,77]. For example, in experiments with propagating squeezed microwaves, phase-locked loops are used to stabilize the relative phase between the pump tones of two JPAs [77–79], ensuring that the interference conditions are maintained over long timescales [79]. With current technology, relative phase fluctuations can be reduced to a small fraction of a radian, which is sufficient to preserve the interference effect and realize the predicted entanglement enhancement.

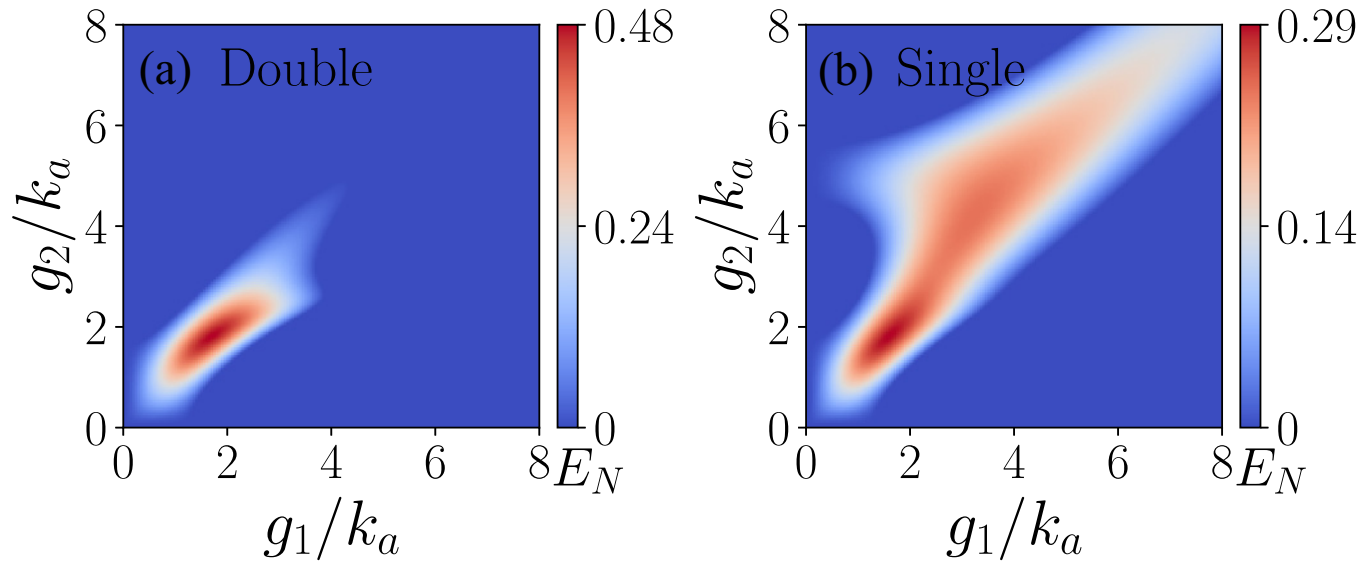

FIG. 7. Effect of the magnon-cavity coupling strengths on the magnon-magnon entanglement in the (a) double- and (b) single-squeezed configurations, where $\Delta_{a_1} = \Delta_{a_2} = -J$ and $\Delta_{m_1} = -\Delta_{m_2} = 0.5J$.

In the present cavity-magnon system, the steady state is reached on timescales determined by the dissipation rates of the cavity and magnon modes, typically on the order of $1/\kappa_a$ and $1/\kappa_m$. For microwave cavity-magnon platforms based on YIG spheres, these rates are usually in the MHz range, corresponding to steady-state timescales from hundreds of nanoseconds to microseconds under continuous driving [6,80]. These timescales are substantially shorter than experimentally demonstrated phase-coherence times in phase-locked squeezed microwave sources [64].

Moreover, although the entanglement enhancement reported here is phase sensitive, it is not an all-or-nothing effect. Finite phase drift primarily reduces the interference contrast and therefore quantitatively decreases the achievable entanglement, rather than eliminating the underlying physical mechanism. This behavior is consistent with previous theoretical and experimental studies on phase-sensitive interference of squeezed microwave fields [76,77].

## IV. CONCLUSION

In summary, we have presented a scheme to generate and enhance remote magnon-magnon entanglement in a coupled cavity-magnon system using only beam-splitter interactions. We show that when a single-squeezed vacuum field drives one cavity mode, quantum correlations can only be transferred to two remote magnon modes through a *determined* supermode formed by the hybridized cavity modes, thereby establishing measurable entanglement between two magnon modes. In the double-squeezed configuration, where each cavity is driven by an independent squeezed field, both supermodes can be *selectively* resonant with the squeezed field by tuning the cavity detunings to transfer quantum correlations to the two magnon modes. In this case, two squeezed fields interfere coherently through the supermode interface, leading to a pronounced *enhancement* of magnon-magnon entanglement. This quantum interference further enables active phase control of the entanglement and markedly improves its robustness against thermal noise. Under experimentally realistic parameters, the entanglement survival temperature increases from approximately 260 mK in the single-squeezed case to about 450 mK in the double-squeezed case. Our work establishes a versatile and controllable approach to enhancing quantum correlations through interference between double-squeezed fields, enabling phase-controlled and robust entanglement engineering in cavity-magnon systems with beam-splitter interactions.

## ACKNOWLEDGMENTS

This work was supported by the Natural Science Foundation of Zhejiang Province (Grants No. LQN26A040020 and No. LY24A040004), the Zhejiang Province Key R&D Program of China (Grant No. 2025C01028), the Shenzhen International Quantum Academy (Grant No. SIQA2024KFKT010), and the Graduate Student Innovation Fund of Wenzhou University (Grant No. 3162025003044).

## DATA AVAILABILITY

The data that support the findings of this article are openly available [81].